\documentclass[11pt,a4paper]{article}
\usepackage{jcappub}
\usepackage[sort&compress]{natbib}
\usepackage{hyperref,ifthen}
\usepackage{epstopdf}
\usepackage{color}
\newcommand{\beq}[1]{\begin{equation}\label{#1}}
 \newcommand{\eeq}{\end{equation}}
 \newcommand{\bea}{\begin{eqnarray}}
 \newcommand{\eea}{\end{eqnarray}}
\begin{document}

\title{Ultra-compact structure in radio quasars as a cosmological probe:
a revised study of the interaction between cosmic dark sectors}

\author{Xiaogang Zheng,$^{1,2}$}
\author{Marek Biesiada,$^{1,2}$}
\author{Shuo Cao,$^1$}
\author{Jingzhao Qi,$^1$}
\author{Zong-Hong Zhu$^1$}

\affiliation{ $^1$ Department of Astronomy, Beijing Normal
University, Beijing 100875, China \\
$^2$ Department of Astrophysics and Cosmology, Institute of Physics,
University of Silesia, Uniwersytecka 4, 40-007 Katowice, Poland\\
}

\abstract{A new compilation of $120$ angular-size/redshift data for
compact radio quasars from very-long-baseline interferometry (VLBI)
surveys motivates us to revisit the interaction between dark energy
and dark matter with these probes reaching high redshifts $z\sim
3.0$. In this paper, we investigate observational constraints on
different phenomenological interacting dark energy (IDE) models with
the intermediate-luminosity radio quasars acting as individual
standard rulers, combined with the newest BAO and CMB observation
from \emph{Planck} results acting as statistical rulers. The results
obtained from the MCMC method and other statistical methods
including Figure of Merit and Information Criteria show that: (1)
Compared with the current standard candle data and standard clock
data, the intermediate-luminosity radio quasar standard rulers ,
probing much higher redshifts, could provide comparable constraints
on different IDE scenarios. (2) The strong degeneracies between the
interaction term and Hubble constant may contribute to alleviate the
tension of $H_0$ between the recent \textit{Planck} and \emph{HST}
measurements. (3) Concerning the ranking of competing dark energy
models, IDE with more free parameters are substantially penalized by
the BIC criterion, which agrees very well with the previous results
derived from other cosmological probes. }

\maketitle

\section{Introduction}
Dark energy has become one of the most important issues of modern
cosmology since many astrophysical and cosmological observations
such as Type Ia Supernovae (SN Ia) \citep{Riess98,Perlmutter99},
Large Scale Structure (LSS) \citep{Eisenstein05} and cosmic
microwave background radiation (CMB) \citep{Komatsu11} coherently
indicate that the universe is undergoing an accelerated expansion at
the present stage. The most simple candidate for the uniformly
distributed material component responsible for this behaviour, is
some form of vacuum energy density or cosmological constant
$\Lambda$. Despite its simplicity, the simple cosmological constant
is always entangled with the well-known coincidence problem: the
matter density $\rho_m$ decreases with the expansion of the universe
as $a^{-3}$ and the density of cosmological constant $\rho_\Lambda$
does not change with the expansion of the universe, whereas the
matter density is comparable with the dark energy density today.
More recent works have implied that this simple $\Lambda$CDM model
encounters other problems. For instance, a tension between
$\Lambda$CDM and the measurements of Hubble parameter ($H(z)$) has
been extensively discussed in the framework of different $Om$
diagnostics \citep{Sahni14,Ding15,Zheng16}. Moreover, the local
measurement of Hubble constant ($H_0$) through the Wide Field Camera
3 (WFC3) on the Hubble Space Telescope (\emph{HST}) \citep{Riess16}
suggests a higher value of $H_0$ than the prediction of $\Lambda
CDM$ model from \emph{Planck} CMB data \citep{Ade16a}.
In order to meet these problems, the consideration of the
possibility of exchanging energy between DE and DM through
an interaction term seems feasible, as the coupling between these dark
sectors can provide a mechanism to alleviate the coincidence problem
\citep{Amendola00,Olivares06,Wang16} and provide
possible a larger value of $H_0$ derived from the recent \emph{HST}
measurement.

However, the nature of DE and DM remains a mystery and most of the
investigation of interaction is based on the phenomenological assumptions.
As extensively considered in the literature (see
e.g.~~\citet{Dalal01, Guo07, Li16}),
we assume that dark energy and dark matter exchange energy through
an interaction term $Q$, while there is no interaction between
baryonic matter and dark energy
\begin{eqnarray}
 &&\dot{\rho}_X+3H\left(\rho_X+p_X\right)=-Q,\nonumber\\
 &&\dot{\rho}_c+3H\rho_c=Q,\nonumber\\
 &&\dot{\rho}_b+3H\rho_b=0.
 \label{eq1}
\end{eqnarray}
where $\rho_{X}$, $\rho_{c}$ and $\rho_{b}$ denote the energy
density of dark energy, cold dark matter and baryonic matter,
respectively. In fact, in the standard model of particle physics
combined with the solar system experiments \citep{Will01}, the
coupling between dark energy and baryons is basically constrained to
irrelevantly small values. This is in agreement with the results
obtained in the modified interacting dark energy (MIDE) model as a
candidate to describe possible interaction between dark energy and
dark matter as well as that between dark energy and baryonic matter
\citep{Cao15a}.

Note that with the ansatz (\ref{eq1}), the total energy density is
conserved: $\dot{\rho}_{tot}+3H\left(\rho_{tot}+p_{tot}\right)=0.$
If $Q$ is a non-zero function of the scale factor, the interaction
makes $\rho_m$ and $\rho_{X}$ to deviate from the standard scaling.
Considering from continuity equations, the interaction term $Q$
must be a function multiplied by the energy density and a quantity
said the inverse of time which can be considered by the Hubble factor $H$.
The simplest assumptions including a freedom of choice
concerning the form of the energy density are $Q=Q(H\rho_{c})$ and
$Q=Q(H\rho_{X})$ (see \citet{Cao13} and references therein). We
assume that, for the phenomenological interaction models, the
equation of state (EoS) of dark energy $w_X\equiv p/\rho$ is
constant in spatially flat FRW metric. The standard cosmology
without interaction between dark energy and dark matter is recovered
with $Q=0$, while $Q\neq0$ denotes non-standard cosmology. On the
other hand, the value of $Q$ determines the extent of interaction
and transfer direction between dark energy and dark matter, i.e.,
$Q<0$ indicates that the energy is transferred from dark matter to
dark energy, while $Q>0$ denotes that the energy is transferred from
dark energy to dark matter.

In the previous works, different interacting dark energy models have
been discussed with various cosmological observations
\citep{Feng08,Cao11,Pan12,Cao13,Xia16,Costa17}. Recently, the angular size
of compact structure in radio quasars versus redshift data from the
very-long-baseline interferometry (VLBI) observations have become an
effective probe in cosmology and astrophysics \citep{Cao17a,Cao17b}.
Based on a 2.29 GHz VLBI all-sky survey of 613 milliarcsecond
ultra-compact radio sources, \citet{Cao17b} extracted a sub-sample
of 120 intermediate-luminosity quasars in the redshift range of
$0.46<z<2.8$, which can be used as cosmological standard rulers with
the linear sizes calibrated by a cosmological-model-independent
method. Compared with the commonly-used SN Ia standard candles
($z\sim1.4$), the advantage of this data set is that quasars are
observed at much higher redshifts ($z\sim 3.0$), which motivates us
to investigate the possible interaction between dark energy and dark
matter at higher redshifts. To reduce the uncertainty and put
tighter constraint on the value of the coupling, we add, in our
discussion, the CMB observation from the \emph{Planck} results as
well as the BAO measurements both from the low-$z$ Galaxy and
higher-$z$ Lyman-$\alpha$ Forests (Ly$\alpha$F) data. We expect that
sensitivities of measurements of different observables can give
complementary results on the coupling between dark sectors.

This paper is organized as follows: The Friedmann equations in the
IDE models are presented in Section~\ref{sec:model}. In
Section~\ref{sec:data}, we introduce the observational data
including QSO, BAO and CMB and the corresponding methodology. In
Section~\ref{sec:result}, we perform a Markov Chain Monto Carlo
(MCMC) analysis, and furthermore apply the Figure of Merit (FoM) and
two model selection techniques, i.e., the Akaike Criterion (AIC) and
Bayesian Information Criterion (BIC). Finally the results are
summarized in Section~\ref{sec:conclu}.

\section{The interacting dark energy models} \label{sec:model}

In this paper, in order to derive stringent constraints on the
interaction between DE and DM, having in mind the well known
degeneracy between model parameters, we assume a flat universe which
is strongly supported by CMB data. In a zero-curvature universe
filled with ordinary pressureless dust matter (cold dark matter plus
baryons), radiation and dark energy, the Friedmann equation reads
\begin{equation}
\rho=\rho_{b}+\rho_{c}+\rho_{X}+\rho_{r}=3H^2(z)/8\pi G
\end{equation}
where $\rho_{r}$ is the energy density of radiation. We remark here
that the effect of this radiation term, which is negligible for
low-$z$ observations, should be taken into account when considering
the redshifts at photo-decoupling epoch $z_*$ and baryon-drag epoch
$z_d$ for the comoving sound horizon in CMB and BAO. Obviously, the
dimensionless expansion rate of the Universe, $E(z)$, without the
interaction between dark matter and dark energy can be expressed as
\begin{equation}
E^2(z; \textbf{p})=
(\Omega_{b0}+\Omega_{c0})(1+z)^3+\Omega_{r0}(1+z)^4+ \Omega_{X}(z).
\end{equation}
where $\Omega_{b0}=(8\pi G\rho_{b0})/(3H^2_0)$ is the current
baryonic matter component, $\Omega_{c0}=(8\pi G\rho_{c0})/(3H^2_0)$
is the current dark matter component, the current radiation
component $\Omega_{r0}=(8\pi G\rho_{r0})/(3H^2_0)=4.1736\times
10^{-5}h^{-2}$ \citep{Komatsu09}, and the dark energy component
\begin{equation}
\Omega_{X}(z)=(1-\Omega_{b0}-\Omega_{c0}-\Omega_{r0})\times(1+z)^{3(1+w)}.
\end{equation}

In our analysis, we assume that dark energy and cold dark matter
exchange energy through interaction term $Q$ proportional to
different energy densities $\rho_{i}$, i.e.,
${Q}=3{\gamma_{i}}{H}{\rho}_{i}$, where $\gamma_{i}$ is a
dimensionless constant. The simplest assumption with the freedom to
choose the form of the energy density are
 \begin{equation}\label{eq5}
  Q_1=3\gamma_cH\rho_c
  \end{equation}
and
  \begin{equation}\label{eq6}
   Q_2=3\gamma_XH\rho_X,
  \end{equation}
where the constants $\gamma_c$ and $\gamma_X$ quantify the extent of
interaction between cold dark matter and dark energy. Correspondingly,
$\gamma_{i}=0$ indicates that there is no interaction between dark
energy and dark matter, while the energy is transferred from dark
matter to dark energy when $\gamma_{i}<0$, and from dark energy to
dark matter when $\gamma_{i}>0$. For the first case, the interaction
term $Q$ is proportional to the energy density of cold dark matter
$\rho_c$. Combining Eq.(\ref{eq1}) and the corresponding expression
of Eq.(\ref{eq5}), we obtain the energy density of dark matter and
dark energy as $\rho_{c}=\rho_{c0}(1+z)^{3(1-\gamma_{c})}$ and
$\rho_{X}=C(1+z)^{3(1+w)}-\frac{\gamma_{c}\rho_{c0}}{\gamma_{c}+w}(1+z)^{3(1-\gamma_{c})}$,
where $C$ is an integral constant to be determined. From the
Friedmann equation
($H^{2}(z)=8{\pi}G(\rho_{b}+\rho_{c}+\rho_{X}+\rho_{r})/3$) combined
with $H_0^2=8{\pi}G\rho_{cr}/3$, the Hubble parameter for this
interacting dark energy model (which is denoted as $\gamma_{c}$IwCDM
model hereafter) is written as
\begin{eqnarray}
{E^2(z)}&=&\frac{w{\Omega}_{c0}}{\gamma_{c}+w}(1+z)^{3(1-\gamma_{c})}+\Omega_{b0}(1+z)^3+\Omega_{r0}(1+z)^4\nonumber\\
        &+&(1-\Omega_{b0}-\Omega_{r0}-\frac{w{\Omega}_{c0}}{\gamma_{c}+w})(1+z)^{3(1+w)} \,.
\end{eqnarray}
It can be seen that vanishing interaction term $\gamma_c=0$,
corresponds to the well-known wCDM parametrization. Moreover, in our
analysis we will consider another special case of the
$\gamma_{c}$IwCDM model with $w=-1$, i.e., the possibility that the
cosmological constant (vacuum energy) exchange energy with dark
matter through an interaction term $\gamma_{c}$, which is denoted as
$\gamma_{c}$I$\Lambda$CDM model hereafter. For the second case, when
the interaction is proportional to the dark energy density $\rho_X$
(which is denoted as $\gamma_{X}$IwCDM model hereafter), the
dimensionless expansion rate of the Universe reads
\begin{eqnarray}
{E^2(z)}&=&\frac{w(\Omega_{b0}+\Omega_{c0})+\gamma_X}{w+\gamma_X}(1+z)^3+\Omega_{r0}(1+z)^4\nonumber\\
        &+&\frac{w(1-\Omega_{b0}-\Omega_{c0})}{w+\gamma_{X}}(1+z)^{3(1+w+\gamma_X)} \,.
\end{eqnarray}
which corresponds to the $\gamma_{X}$I$\Lambda$CDM model with
$w=-1$.

Let us emphasize the use of the Hubble constant $H_0$ in our
analysis. Considering the tension of its precise value between
different astrophysical methods, i.e., Cepheids from the Hubble
Space Telescope (\emph{HST}) key project \citep{Riess16} and CMB
data from the \emph{Planck} satellite experiment \citep{Ade16b}, we
will keep the Hubble constant as a free parameter, which will be
constrained by the observational data described below. Therefore,
the interacting dark energy models considered in this paper have
five parameters ($\Omega_{b0}$, $\Omega_{c0}$, $w$, $\gamma_{i}$ and
$H_0$), where $\Omega_{b0}$ and $\Omega_{c0}$ specify the current
density of baryonic matter and dark matter, $w$ denotes the equation
of state of dark energy, $\gamma_{c}$ and $\gamma_{X}$ presents how
strongly dark energy interacts with dark matter.

\section{Observational data and methodology}\label{sec:data}

In this work, we will consider a combination of three types of
standard rulers to derive the information of angular diameter
distances and thus the interaction between dark sectors at different
redshifts, i.e., the compact radio quasars data from VLBI, the
cosmic microwave background (CMB) measurements from the
\emph{Planck} results, and the baryonic acoustic oscillations (BAO)
from the low-$z$ galaxy data and the higher-$z$ Lyman-$\alpha$
Forests (Ly$\alpha$F) data. The first probes can be considered as
individual standard rulers while the other two probes are treated as
statistical standard rulers in cosmology. For comparison, we also
considered two more standard probes alternative to standard rulers:
standard candles -- type Ia supernovae (SN Ia) and standard
chronometers -- passively evolving galaxies (more precisely, $H(z)$
data derived from them).

\subsection{Individual standard rulers: QSO}

It is well known that for objects with determined intrinsic physical
size, one can determine the corresponding angular diameter distances
by measuring their angular sizes at different redshifts.
In the case of compact radio-sources used here,
characteristic angular size obtained from VLBI has been obtained as
\begin{equation}
\theta=\frac{2\sqrt{-ln{\Gamma}ln{2}}}{\pi{B}}
\end{equation}
where $B$ is the interferometer baseline and the visibility modulus
$\Gamma$ is the ratio of the total flux density $S_{tot}$ and
correlated flux density $S_{corr} $ \citep{Preston85}. We
note that all quantities derived from the $S_{corr}/S_{tot}$ ratio
(including the apparent angular size) are weakly dependent on the
source orientation, considering the well-known fact that for most of
the strongly core dominated radio sources the viewing angle should
be close to the $1/\gamma_{jet}$ value. Angular sizes of the
compact structure $\theta_{obs}$ and their corresponding
uncertainties $\sigma_{\theta}(z_i)$ (comprising both statistical
and systematical uncertainty) can be found in \citep{Cao17a}. As is
extensively discussed in the literature \citep{Gurvits94}, the
linear sizes of compact structure in radio sources could depend on
the luminosity and redshift as
\begin{equation}
l_m=lL^{\beta}(1+z)^{n}
\end{equation}
where $\beta$ and $n$ are the two parameters respectively
characterizing the ``linear size - luminosity'' and ``linear size -
redshift'' relations. Let us recall that besides cosmological
evolution of the linear size with redshift, the parameter n may also
characterize the dependence of the linear size on the emitted
frequency, as well as image blurring due to scattering in the
propagation medium.

The possibility to alleviate the dependence of $l_m$ on the source
luminosity and redshift were discussed early by
\citet{Gurvits99,Vishwakarma01}. Then \citet{Cao15b} using a
compilation of mixed population of radio sources including different
optical counterparts found they can not act as standard ruler. Most
recently, the analysis of \citep{Cao17a} indicated that only a
sub-sample of intermediate-luminosity radio quasars ($10^{27}$ W/Hz
$<L<10^{28}$ W/Hz) displayed negligible dependence on luminosity and
redshift ($\beta\simeq10^{-4},|n|\simeq10^{-3}$) and therefore could
serve as a standard ruler in cosmology. Let us stress that sample
selection in terms of luminosity, which involves the knowledge of
angular diameter distances to the quasars, can be performed in a
robust way not depending on the details of cosmological model (for
details see \citep{Cao17a}) and does not induce circularity problems
in using QSOs for cosmological model parameter inference.
\citet{Cao17b} also confirmed that the uncertainties of $\beta$ and
$n$ do not influence the best-fit cosmological parameters
noticeably. For the observational quasar data, we adopt the
measurements of 120 intermediate-luminosity radio quasars in the
redshift range $0.46<z<2.80$. In previous works, the
intrinsic linear size $l_m$ was calibrated using the angular
diameter distances estimated from two different methods: the supernova
distance modulus \citep{Cao17a} and the cosmic chronometers (themselves being cosmology-independent probe) using
a robust reconstruction technique called the Gaussian process
\citep{Cao17b}. In this paper, we choose to use the the linear size
of this standard ruler calibrated as $l_m = 11.03\pm0.25$ pc through
the cosmological-model-independent method \citep{Cao17b}. The
angular size of the compact structure in intermediate-luminosity
radio quasars, observed at redshifts $z$, can be theoretically
expressed as
\begin{equation}
\theta_{th}(z)=\frac{l_m}{D_A(z)}
\end{equation}
where $D_{A}$ is the angular diameter distance computed in the
models of interest, which reads
\begin{equation}
D_{A}(z;{\bf{p}})=\frac{c}{H_0}\frac{1}{1+z}\int^{z}_{0}\frac{dz'}{E(z';\bf{p})}
\end{equation}
in flat Friedman-Robertson-Walker metric.

In order to constrain the model
parameters $\bf{p}$ using these quasar data, we define the
likelihood function $\mathcal{L}_{QSO}\propto\exp[-\chi_{QSO}^2(z;
\textbf{p})/2]$, where $\chi_{QSO}^2$ is related to the quasar
sample as
\begin{equation}
\chi^{2}_{QSO}=\Sigma^{120}_{i=1}\frac{[\theta_{obs}(z_{i})-\theta_{th}(z_{i};{\bf{p}})]^{2}}{\sigma_{\theta}^{2}(z_{i})}.
\end{equation}

\subsection{Statistical standard rulers: CMB and BAO}

The first statistical standard ruler used in our analysis is the
cosmic microwave background (CMB), providing the sound horizon scale
at high redshift ($z\sim 1089$), useful for determining the
properties of dark sectors in the Universe. For the CMB data, we
will use the measurements of several quantities derived from
$Planck$ \citep{Ade16b}, which include the acoustic scale ($l_A$),
the shift parameter ($R$), and the baryonic matter fraction at the
redshift of recombination ($\Omega_{b0}h^2$). Firstly, the acoustic
scale is expressed as
\begin{equation}
l_A\equiv(1+z_*)\frac{\pi D_A(z_*)}{r_s(z_*)}
\end{equation}
where the redshift of photon-decoupling period can be calculated as
\citep{Hu96}
\begin{equation}
z_*=1048[1+0.00124(\Omega_{b0}h^2)^{-0.738}][1+g_1(\Omega_{m0}h^2)^{g_2}]
\end{equation}
\begin{equation}
g_1=\frac{0.0783(\Omega_{b0}h^2)^{-0.238}}{1+39.5(\Omega_{b0}h^2)^{0.763}},
g_2=\frac{0.560}{1+21.1(\Omega_{b0}h^2)^{1.81}}
\end{equation}
The comoving sound horizon can be parameterized as
\begin{eqnarray}
r_s(z_*)&=&\int^t_0\frac{c_sdt'}{a}=\frac{c}{H_0}\int^{\infty}_{z_*}\frac{c_sdz}{E(z)} \nonumber\\
        &=&\frac{c}{H_0}\int^{a_*}_0\frac{da}{a^2E(a)\sqrt{3(1+\overline{R_{b}}a)}} \,.
\end{eqnarray}
with $\overline{R_{b}}=31500(T_{CMB}/2.7K)^{-4}\Omega_{b0}h^2$ and
$T_{CMB}=2.7255K$. Note that the current radiation component is
related to the matter density as
$\Omega_{r0}=\Omega_{m0}/(1+z_{eq})$, where
$z_{eq}=2.5\times10^4\Omega_{m0}h^2(T_{CMB}/2.7K)^{-4}$. Secondly,
the other commonly-used CMB shift parameter reads \citep{Bond97}
\begin{equation}
R(z_*)\equiv\frac{(1+z_*)D_A(z_*)\sqrt{\Omega_{m0}H_0^2}}{c}
\end{equation}
Model parameters can be estimated from the CMB data using the
likelihood based on the $\chi^2$ function defined as
\begin{equation}
\chi^2_{CMB}=\Delta{P}^T_{CMB}C^{-1}_{CMB}\Delta{P}_{CMB}
\end{equation}
where $\Delta{P}_{CMB}$ is the difference between the theoretical
distance priors and the observational counterparts. The
corresponding inverse covariance matrix $C^{-1}_{CMB}$ can be found
in Table.~4 provided by \citet{Ade16b}.

The second statistical standard ruler applied in our analysis is the
baryonic acoustic oscillations (BAO) scale, the measurements of
which are derived from both the low-$z$ galaxy and higher-$z$
Lyman-$\alpha$ Forests(Ly$\alpha$F) data. For the lower-redshift BAO
observations, we turn to the latest measurements of acoustic-scale
distance ratio from the 6-degree Field Galaxy Survey (6dFGS)
\citep{Beutler11}, the 'main galaxy sample' from Sloan Digital Sky
Survey (SDSS-MGS) \citep{Ross15}, and the two principal Baryon
Oscillation Spectroscopic Survey (BOSS) galaxy samples (BOSS-LOWZ
and BOSS-CMASS) \citep{Anderson14}, while for the the higher-$z$
measurement is derived from SDSS DR12 \citep{Bautista17}. Now three
different types of distance ratios $D_V(z)/r_s(z_d)$,
$D_M(z)/r_s(z_d)$ and $D_H(z)/r_s(z_d)$ are listed in
Table.~\ref{tableBAO}. We remark that $r_s(z_{d})$ is the comoving
sound horizon at the baryon-drag epoch $z_{d}$ \citep{Eisenstein98}
\begin{equation}
z_d=\frac{1291(\Omega_{m0}h^2)^{0.251}}{1+0.659(\Omega_{m0}h^2)^{0.828}}[1+b_1(\Omega_{b0}h^2)^{b_2}]
\end{equation}
with
\begin{eqnarray}
&&b_1=0.313(\Omega_{m0}h^2)^{-0.419}[1+0.607(\Omega_{m0}h^2)^{0.674}), \nonumber\\
&&b_2=0.238(\Omega_{m0}h^2)^{0.223}.
\end{eqnarray}
$D_V(z)$, $D_M(z)$ and $D_H(z)$ represent the volume-averaged
effective distance ($[(1+z)^2D_A^2(z)cz/H(z)]^{1/3}$), the comoving
angular-diameter distance ($(1+z)D_A(z)$), and the Hubble distance
($c/H(z)$), respectively. The $\chi^2$ function constructed from the
BAO observations is denoted as $\chi^2_{BAO}$ hereafter
and it can be expressed as
\begin{equation}
\chi^2_{BAO}=\sum^n_{i=1}\frac{(\mathcal{R}_{th}-\mathcal{R}_{obs})^2}{(\Delta \mathcal{R}_{obs})^2}
\end{equation}
where $\mathcal{R}$ stands for three different types of distance
ratios mentioned above.

\begin{table*}
\centering \caption{Distance ratios from recent BAO measurements.}
{{\scriptsize
 \begin{tabular}{c c c c c c} \hline\hline
   Survey  & $z$ & $D_V(z)/r_s(z_d)$ & $D_M(z)/r_s(z_d)$ & $D_H(z)/r_s(z_d)$ & Reference \\ \cline{1-6}
6dFGS  & $0.106$ & $3.047\pm0.137$ & $-$ & $-$ & \citet{Beutler11} \\
SDSS-MGS & $0.15$ & $4.480\pm0.168$ & $-$ & $-$ & \citet{Ross15} \\
BOSS-LOWZ & $0.32$ & $8.594\pm0.095$ & $8.774\pm0.142$ & $25.89\pm0.76$ & \citet{Anderson14} \\
BOSS-CMASS & $0.57$ & $13.757\pm0.142$ & $14.745\pm0.237$ &
$21.02\pm0.52$ & \citet{Anderson14} \\ \cline{1-6}
SDSS DR12 & $2.33$ & $-$ & $37.77\pm2.73$ & $9.07\pm0.31$ & \citet{Bautista17}\\
 \hline \hline
\end{tabular}} \label{tableBAO}}
\end{table*}

\subsection{Alternative standard probes for comparison: SN Ia and $H(z)$}

As it is well known, the evidence for cosmic
acceleration came first from other type of standard probes in
cosmology, i.e., type Ia supernovae (SN Ia) probing the
luminosity distance $D_L$. Later on, the ages of passively
evolving early-type galaxies (cosmic chronometers) became
available providing direct measurements of the Hubble parameter
$H(z)$ at different redshifts. In order to compare our QSO
fits, with the results obtained using these alternative probes,
we also considered the latest Union2.1 compilation \citep{Suzuki12}
consisting of 580 SN Ia in the redshift range $0.014 < z < 1.415$
and 30 Hubble parameter measurements \citep{Zheng16} in the
redshift range $0.07<z<1.965$ obtained from cosmic chronometers.

The $\chi^2_{SN}$ of the Union2.1 SN Ia is given by
\begin{equation}
\chi^2_{SN}=(\mu_{obs}-\mu_{th})^TC_{SN}^{-1}(\mu_{obs}-\mu_{th})
\end{equation}
where $\mu=5log_{10}(D_L/Mpc)+25$ is the distance modulus and
$C_{SN}$ is the covariance matrix, and the $\chi^2_{H(z)}$ of
the $H(z)$ data is given by
\begin{equation}
\chi^2_{H(z)}=\Sigma^{30}_{i=1}\frac{[H_{obs}(z_{i})-H_{th}(z_{i};\bf{p})]^{2}}{\sigma^{2}_{H_{obs}(z_i)}}
\end{equation}
where $H_{obs}(z_{i})$ and $\sigma_{H_{obs}(z_i)}$ are the $30$
Hubble parameter measurements and their uncertainties, respectively.

In the next section we will present the results of
joint analysis involving the above mentioned probes in different
combinations based on minimizing the following chi-square functions:
$\chi^2_{C1}=\chi^2_{CMB}+\chi^2_{BAO}$,
$\chi^2_{C2}=\chi^2_{QSO}+\chi^2_{CMB}+\chi^2_{BAO}$,
$\chi^2_{C3}=\chi^2_{CMB}+\chi^2_{BAO}+\chi^2_{SN}+\chi^2_{H(z)}$, and
$\chi^2_{C4}=\chi^2_{QSO}+\chi^2_{CMB}+\chi^2_{BAO}+\chi^2_{SN}+\chi^2_{H(z)}$.
In adopting the MCMC approach, we generate a chain of sample points
distributed in the parameter space according to the posterior
probability by using the Metropolis-Hastings algorithm with uniform
prior probability distribution, and then repeat this process until
the established convergence accuracy can be satisfied. Our code is
based on CosmoMC \citep{Lewis02} and we generated eight chains after
setting $R-1 = 0.001$ to guarantee the accuracy of the fits.

 \begin{table*}[htp]
\centering \caption{ The marginalized 1$\sigma$ uncertainties of the
parameters $\Omega_{b0}$, $\Omega_{c0}$, $w$, $\gamma_i$, and $H_0$
for different interacting dark energy scenarios, as well as
$\chi^2/d.o.f$ and $FoM$, obtained from the combinations of the data
sets C1 (CMB+BAO), C2 (QSO+CMB+BAO), C3 (CMB+BAO+SN+$H(z)$) and C4 (QSO+CMB+BAO+SN+$H(z)$),
respectively. Corresponding results for the $\Lambda$CDM and wCDM
models are also added for comparison. }
{{\scriptsize
 \begin{tabular}{l c c c c c c c} \hline\hline
$\Lambda CDM$ & $\Omega_{b0}$ & $\Omega_{c0}$ & $w$ & $\gamma$ & $H_0$ & $\chi^2/d.o.f$ & $FoM$ \\ \cline{2-8} \\
 $C1$ & $0.0501^{+0.0011}_{-0.0010}$ & $0.277^{+0.011}_{-0.011}$ & $-1$ & $0$ & $66.48^{+0.81}_{-0.81}$ & $10.35/10$   & $7.33\times10^6$ \\
 $C2$ & $0.0500^{+0.0010}_{-0.0009}$ & $0.276^{+0.010}_{-0.010}$ & $-1$ & $0$ & $66.59^{+0.80}_{-0.80}$ & $364.55/130$ & $7.37\times10^6$ \\
 $C3$ & $0.0499^{+0.0009}_{-0.0009}$ & $0.275^{+0.008}_{-0.008}$ & $-1$ & $0$ & $66.62^{+0.79}_{-0.79}$ & $571.26/620$ & $7.91\times10^6$ \\
 $C4$ & $0.0499^{+0.0009}_{-0.0008}$ & $0.275^{+0.008}_{-0.008}$ & $-1$ & $0$ & $66.66^{+0.79}_{-0.78}$ & $925.36/740$ & $7.96\times10^6$ \\ \\
 \hline\hline
$\gamma_c I\Lambda CDM$ & $\Omega_{b0}$ & $\Omega_{c0}$ & $w$ & $\gamma_c$ & $H_0$ & $\chi^2/d.o.f$ & $FoM$ \\ \cline{2-8} \\
 $C1$ & $0.0487^{+0.0018}_{-0.0015}$ & $0.277^{+0.009}_{-0.009}$ & $-1$ & $-0.0025^{+0.0027}_{-0.0022}$ & $67.75^{+1.25}_{-1.50}$ & $8.95/9$ & $2.95\times10^9$ \\
 $C2$ & $0.0483^{+0.0015}_{-0.0013}$ & $0.276^{+0.009}_{-0.009}$ & $-1$ & $-0.0029^{+0.0021}_{-0.0018}$ & $68.04^{+1.01}_{-1.23}$ & $361.19/129$ & $4.73\times10^9$ \\
 $C3$ & $0.0486^{+0.0014}_{-0.0013}$ & $0.276^{+0.009}_{-0.010}$ & $-1$ & $-0.0025^{+0.0020}_{-0.0019}$ & $67.38^{+0.98}_{-1.06}$ & $569.12/619$ & $4.87\times10^9$ \\
 $C4$ & $0.0485^{+0.0014}_{-0.0012}$ & $0.276^{+0.009}_{-0.010}$ & $-1$ & $-0.0026^{+0.0018}_{-0.0018}$ & $67.88^{+0.92}_{-1.03}$ & $921.38/739$ & $5.51\times10^9$ \\ \\
 \hline\hline

  $\gamma_X I\Lambda CDM$ & $\Omega_{b0}$ & $\Omega_{c0}$ & $w$ & $\gamma_X$ & $H_0$ & $\chi^2/d.o.f$ & $FoM$ \\ \cline{2-8} \\
 $C1$ & $0.0504^{+0.0011}_{-0.0011}$ & $0.276^{+0.010}_{-0.013}$ & $-1$ & $-0.0052^{+0.0077}_{-0.0083}$ & $66.45^{+1.04}_{-0.71}$ & $9.16/9$ & $1.44\times10^9$ \\
 $C2$ & $0.0504^{+0.0010}_{-0.0010}$ & $0.275^{+0.010}_{-0.010}$ & $-1$ & $-0.0052^{+0.0077}_{-0.0083}$ & $66.53^{+0.81}_{-0.70}$ & $363.24/129$ & $1.59\times10^9$ \\
 $C3$ & $0.0503^{+0.0010}_{-0.0010}$ & $0.274^{+0.010}_{-0.010}$ & $-1$ & $-0.0052^{+0.0077}_{-0.0082}$ & $66.55^{+0.77}_{-0.71}$ & $570.04/620$ & $1.63\times10^9$ \\
 $C4$ & $0.0503^{+0.0010}_{-0.0010}$ & $0.274^{+0.009}_{-0.010}$ & $-1$ & $-0.0057^{+0.0077}_{-0.0082}$ & $66.64^{+0.77}_{-0.70}$ & $924.07/739$ & $1.64\times10^9$ \\ \\
 \hline \hline

 $wCDM$ & $\Omega_{b0}$ & $\Omega_{c0}$ & $w$ & $\gamma$ & $H_0$ & $\chi^2/d.o.f$ & $FoM$ \\ \cline{2-8} \\
   $C1$ & $0.0517^{+0.0035}_{-0.0024}$ & $0.282^{+0.013}_{-0.010}$ & $-0.94^{+0.10}_{-0.08}$ & $0$ & $65.61^{+1.28}_{-2.03}$ & $8.86/9$ & $1.01\times10^8$ \\
 $C2$ & $0.0515^{+0.0034}_{-0.0024}$ & $0.281^{+0.012}_{-0.011}$ & $-0.95^{+0.10}_{-0.07}$ & $0$ & $65.83^{+1.24}_{-2.01}$ & $363.29/129$ & $1.05\times10^8$ \\
 $C3$ & $0.0502^{+0.0029}_{-0.0020}$ & $0.275^{+0.012}_{-0.011}$ & $-0.98^{+0.09}_{-0.08}$ & $0$ & $66.54^{+1.45}_{-1.53}$ & $570.96/619$ & $1.28\times10^8$ \\
 $C4$ & $0.0500^{+0.0029}_{-0.0020}$ & $0.275^{+0.012}_{-0.010}$ & $-0.99^{+0.09}_{-0.07}$ & $0$ & $66.64^{+1.23}_{-1.73}$ & $925.13/739$ & $1.31\times10^8$ \\ \\
 \hline \hline

 $\gamma_c IwCDM$ & $\Omega_{b0}$ & $\Omega_{c0}$ & $w$ & $\gamma_c$ & $H_0$ & $\chi^2/d.o.f$ & $FoM$ \\ \cline{2-8} \\
$C1$ & $0.0509^{+0.0069}_{-0.0066}$ & $0.282^{+0.015}_{-0.016}$ & $-0.95^{+0.14}_{-0.15}$ & $-0.0008^{+0.0044}_{-0.0042}$ & $65.96^{+4.69}_{-4.06}$ & $8.76/8$ & $1.26\times10^{10}$ \\
$C2$ & $0.0483^{+0.0045}_{-0.0040}$ & $0.277^{+0.012}_{-0.012}$ & $-1.00^{+0.11}_{-0.11}$ & $-0.0026^{+0.0026}_{-0.0023}$ & $67.96^{+2.79}_{-2.89}$ & $361.19/128$ & $4.50\times10^{10}$ \\
$C3$ & $0.0481^{+0.0032}_{-0.0033}$ & $0.274^{+0.011}_{-0.010}$ & $-1.02^{+0.08}_{-0.09}$ & $-0.0025^{+0.0025}_{-0.0024}$ & $68.26^{+2.62}_{-2.30}$ & $569.07/618$ & $5.50\times10^{10}$ \\
$C4$ & $0.0476^{+0.0029}_{-0.0026}$ & $0.274^{+0.010}_{-0.010}$ & $-1.03^{+0.08}_{-0.07}$ & $-0.0029^{+0.0019}_{-0.0017}$ & $68.62^{+1.71}_{-2.05}$ & $921.16/738$ & $8.88\times10^{10}$ \\ \\
 \hline \hline

 $\gamma_X IwCDM$ & $\Omega_{b0}$ & $\Omega_{c0}$ & $w$ & $\gamma_X$ & $H_0$ & $\chi^2/d.o.f$ & $FoM$ \\ \cline{2-8} \\
 $C1$ & $0.0514^{+0.0050}_{-0.0041}$ & $0.281^{+0.020}_{-0.017}$ & $-0.96^{+0.18}_{-0.14}$ & $-0.0022^{+0.0135}_{-0.0120}$ & $65.83^{+2.46}_{-3.14}$ & $8.85/8$ & $6.26\times10^9$ \\
 $C2$ & $0.0510^{+0.0041}_{-0.0040}$ & $0.278^{+0.017}_{-0.016}$ & $-0.97^{+0.14}_{-0.15}$ & $-0.0042^{+0.0127}_{-0.0109}$ & $66.06^{+2.60}_{-2.58}$ & $363.13/128$ & $8.08\times10^9$ \\
 $C3$ & $0.0498^{+0.0030}_{-0.0028}$ & $0.274^{+0.014}_{-0.013}$ & $-1.03^{+0.12}_{-0.11}$ & $-0.0081^{+0.0122}_{-0.0095}$ & $66.80^{+1.87}_{-1.88}$ & $569.97/618$ & $1.64\times10^{10}$ \\
 $C4$ & $0.0499^{+0.0028}_{-0.0030}$ & $0.273^{+0.013}_{-0.014}$ & $-1.02^{+0.11}_{-0.11}$ & $-0.0070^{+0.0109}_{-0.0097}$ & $66.79^{+2.01}_{-1.73}$ & $923.94/738$ & $1.64\times10^{10}$ \\ \\
  \hline \hline
\end{tabular}} \label{tablea}}
\end{table*}

\section{Results and discussion}\label{sec:result}

\begin{center}
 \begin{figure}[htbp]
\centering
 \includegraphics[angle=0,width=75mm]{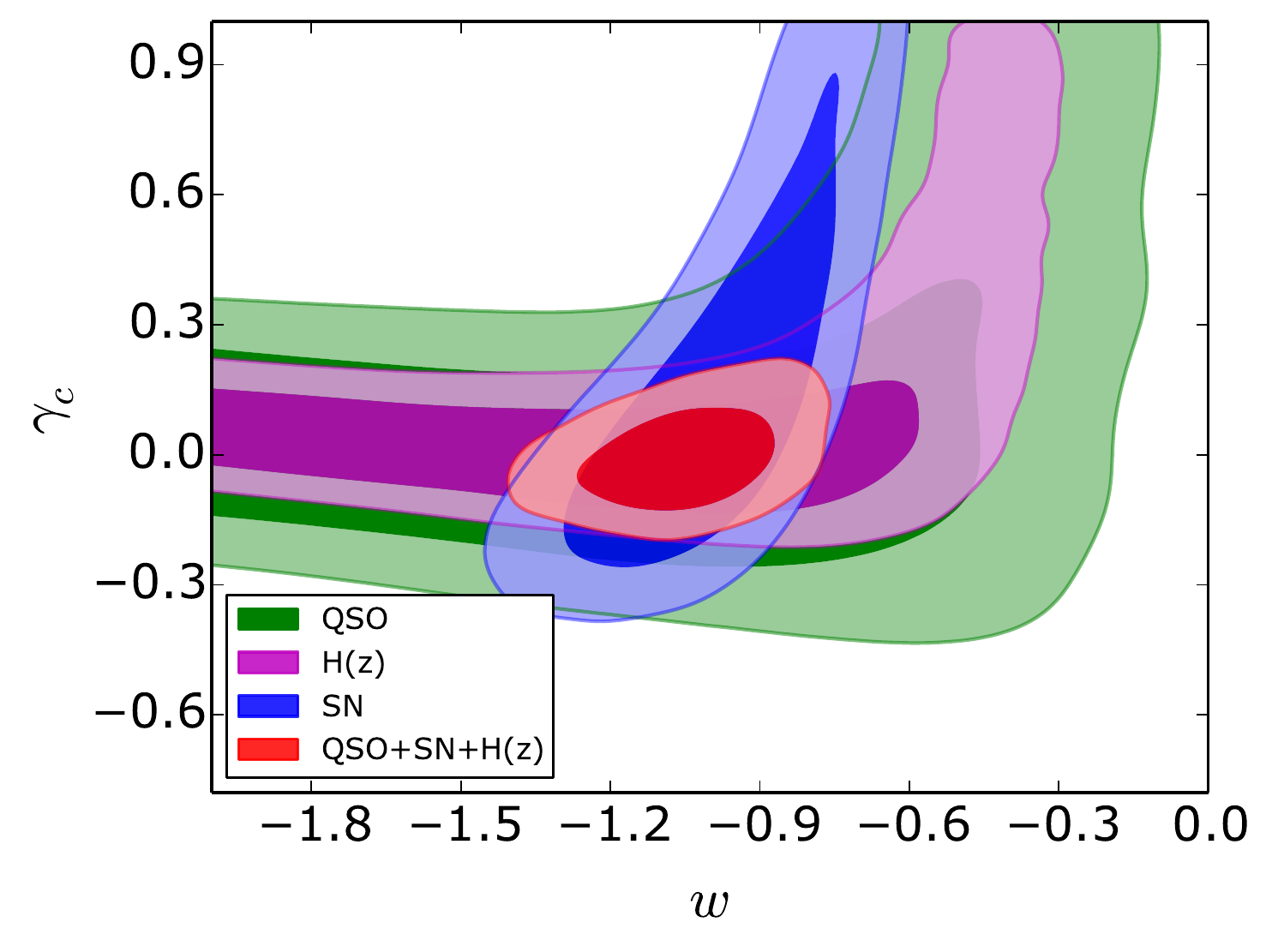}
 \includegraphics[angle=0,width=75mm]{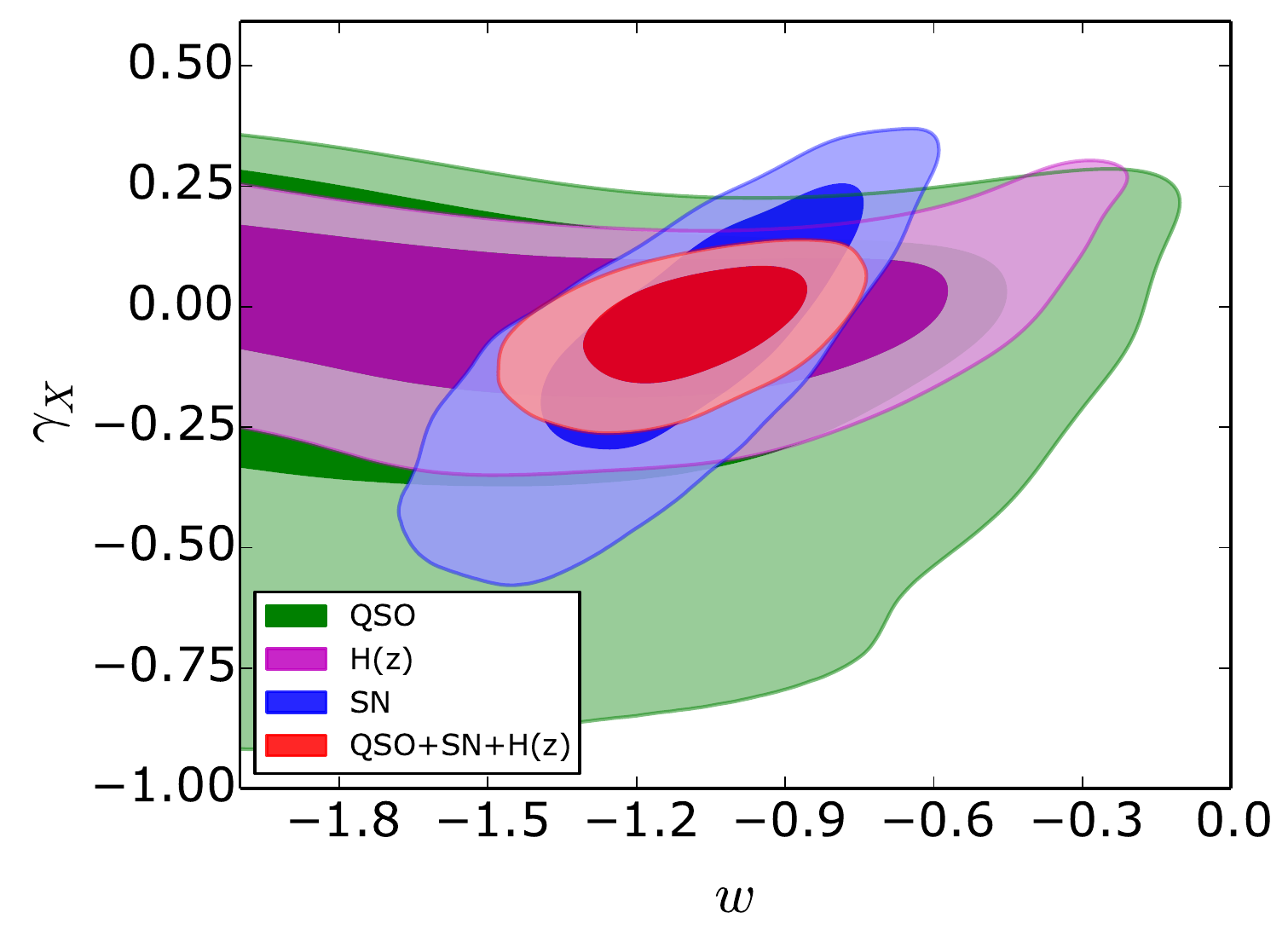}
\caption{\label{figwr} 1$\sigma$ and 2$\sigma$ confidence regions
in the ($w$, $\gamma_i$) for the $\gamma_c$IwCDM model (left panel)
and $\gamma_X$IwCDM model (right panel), which are derived from
three different types of standard probes (QSO, SN and $H(z)$).}
 \end{figure}
 \end{center}

\begin{center}
 \begin{figure}[htbp]
\centering
 \includegraphics[angle=0,width=75mm]{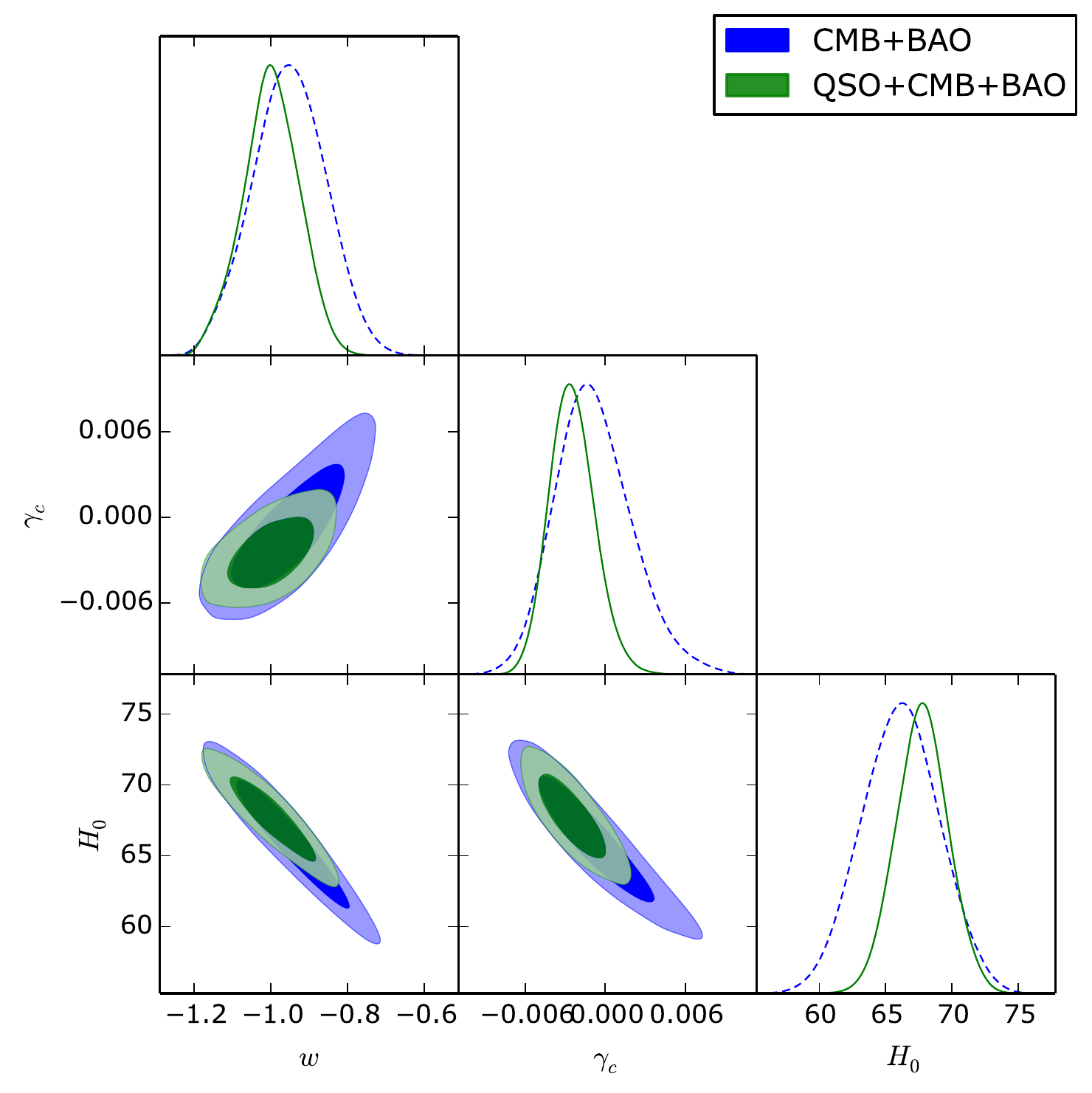}
 \includegraphics[angle=0,width=75mm]{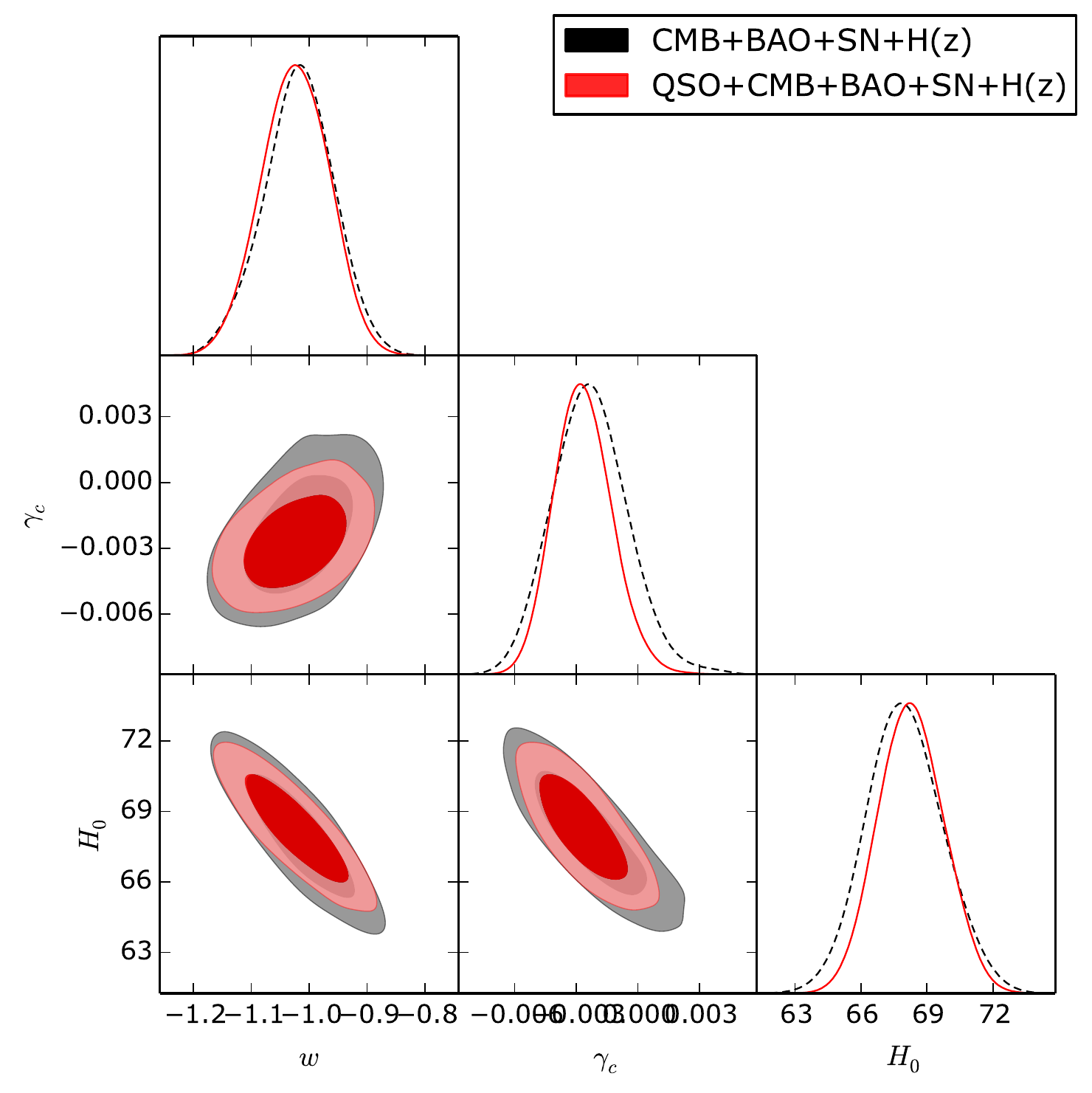}
 \caption{\label{figmodelrc} The 2-D plots and 1-D marginalized
 distributions with 1-$\sigma$ and 2-$\sigma$ contours of the
 $\gamma_c$IwCDM model parameters ($w$, $\gamma_{c}$,
 and $H_0$) obtained with the statistical standard rulers (CMB+BAO),
 combined standard rulers (QSO+CMB+BAO), combination of the standard
 probes (CMB+BAO+SN+H(z)), and all probes including quasars
 (QSO+CMB+BAO+SN+H(z)).}
 \end{figure}
 \end{center}

\begin{center}
 \begin{figure}[htbp]
\centering
 \includegraphics[angle=0,width=75mm]{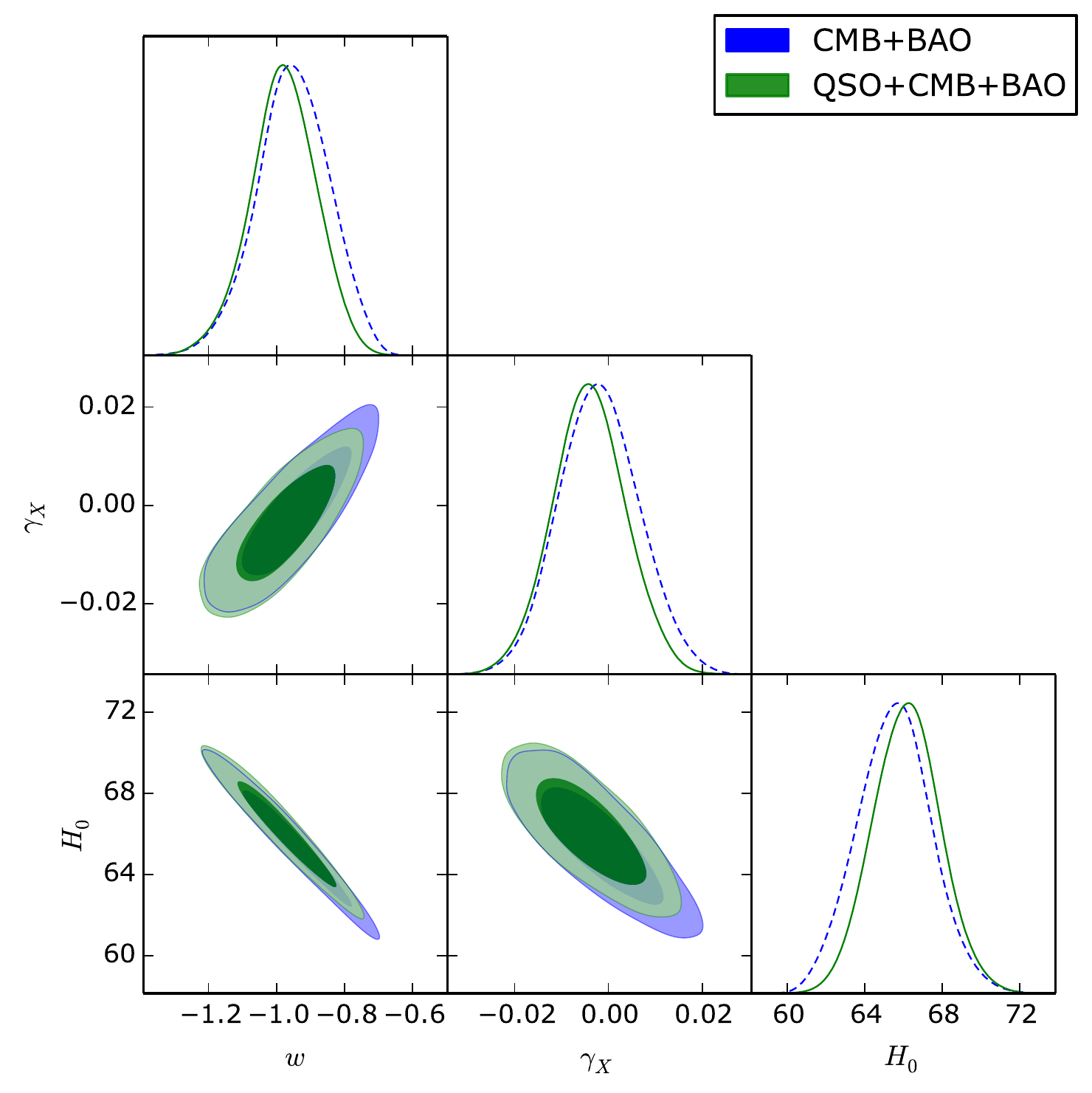}
 \includegraphics[angle=0,width=75mm]{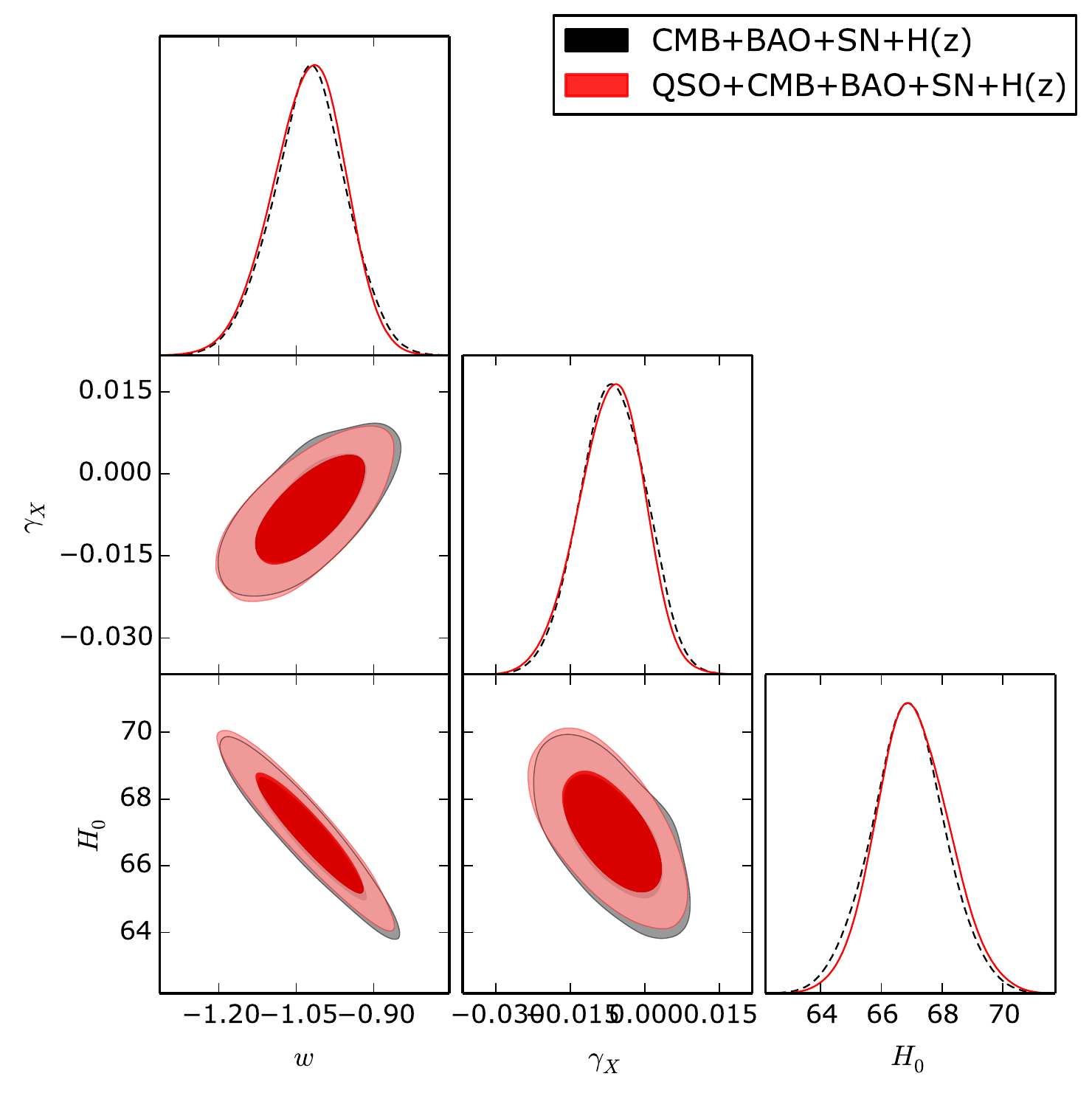}
 \caption{\label{figmodelrd} The same as Fig.~2, but for the $\gamma_X$IwCDM
 model.}
 \end{figure}
 \end{center}

In order to demonstrate the constraining power of quasars on the
interaction between DE and DM, both individual and joint
constraint results will be presented in the following analysis.
Firstly, the fitting results from the compact structure measurements
of 120 quasars are shown in Fig.~\ref{figwr}, in the framework of
the $\gamma_c$ IwCDM and $\gamma_X$IwCDM models. For comparison,
confidence contours concerning the constraints obtained with QSO, SN
and $H(z)$ individually and in combination are displayed as well.
Note that we are focusing on the interaction term between
cosmic dark sectors and individual standard probes cannot tightly
constrain the matter density parameters. Therefore, we have assumed the best
fitted $\Omega_{b0}$ and $\Omega_{c0}$ obtained by the recent
\textit{Planck} observations \citep{Ade16a} as priors. It is clear
that the quasar data could provide constraints comparable to the
other two types of standard probes, i.e., standard candle (SN Ia)
and $H(z)$ from cosmic chronometers. Moreover, our results displayed
in Fig~\ref{figwr} suggest that larger and more accurate sample of
the quasar data can be a valuable complementary probe to test the
properties of dark energy and break the degeneracy in the IDE model
parameters. This tendency will also be preserved in more stringent
constraints obtained from the combination of those three probes.
Secondly, we compared the performance of statistical standard rulers (CMB+BAO) themselves and combined with other probes (CMB+BAO+SN+H(z)) in two settings: with and without inclusion of QSO. Unlike the former case, density parameters $\Omega_{b0}$ and $\Omega_{c0}$ could now be fitted reliably and were treated as a free parameters.
In order to illustrate the comparison of results between different data sets, we
just show the 2-D plots and 1-D marginalized distributions with
1-$\sigma$ and 2-$\sigma$ contours of the parameters ($w$, $\gamma$,
and $H_0$) in Figs.~\ref{figmodelrc}-\ref{figmodelrd}. Corresponding
numerical values of central fits and 1$\sigma$ uncertainties on all
five free parameters can found in Table.~\ref{tablea}.

In the first case of $\gamma_{c}$IwCDM model, in order to illustrate
the performance of the quasar data compare with the statistical
rulers, constraints from statistical standard rulers (CMB+BAO) and
the joint constraint enriched with quasars (QSO+CMB+BAO) are given
in the left panel of Fig.~\ref{figmodelrc}. The right panel of
Fig.~\ref{figmodelrc} shows the confidence contours of model
parameters constrained with the full combination of three types of
standard probes (QSO+CMB+BAO+SN+$H(z)$) and the combination
excluding QSO data (CMB+BAO+SN+$H(z)$). One can clearly see that the
currently compiled quasar data improves the constraints on model
parameters significantly. From the above results, the parameter
$\gamma_c$ capturing the interaction between DE and DM seems to be
vanishing or slightly smaller than 0, which has been noticed by
using the 182 Gold SNIa together with CMB and large-scale structure
for the interacting holographic DE model \citep{Feng07} and by using
the revised Hubble parameter data together with CMB and BAO
\citep{Cao13}. The sample of 59 high redshift calibrated Gamma-Ray
Burst (GRB) data, whose redshift region is more comparable to our
quasar data, combined with BAO observation from SDSS and CMB from
the 7-Year Wilkinson Microwave Anisotropy Probe (WMAP7) also support
this result \citep{Pan12}.

Concerning the $\gamma_{X}$IwCDM model when the interaction is
proportional to the dark energy density $\rho_X$, comparison of the
constraints obtained with statistical standard rulers and combined
standard ruler including quasars indicate that quasar data can
improve the final result noticeably. However, when SN Ia and $H(z)$
data are included the influence of quasars is very small. This can
change when bigger samples of quasar are available in the future.
From the fitting results shown in Fig.~\ref{figmodelrd} and
Table.~\ref{tablea}, the joint analysis provides a small negative
coupling which agrees with the results by using the Gold SNIa
together with CMB and large-scale structure for other interacting DE
models describing the interaction in proportional to the DM energy
density \citep{Guo07}. In addition, the constraining results in this
work with the joint observational data including quasars are more
stringent than previous results for constraining $\gamma_X$IDE model
parameters with other combined observations arising from the 182
Gold SNe Ia samples, the shift parameter of CMB given by the WMAP3
observations, the BAO measurement from the Sloan Digital Sky Survey,
the age estimates of 35 galaxies \citep{Feng08} and the 13 $H(z)$
measurements from the differential ages of red-envelope galaxies as
well as the BAO peaks \citep{Cao13}. Similar conclusions could also
be obtained from the $\gamma_{X}$I$\Lambda$CDM model with $w=-1$,
with the estimation of these cosmic parameters briefly summarized in
Table.~\ref{tablea}.

Now it is worthwhile to make some comments on the results obtained
above. Firstly, let us note that, following \citet{Cao17b}, we added
extra 10\% uncertainties to the observed angular sizes.
This additional uncertainty is equivalent to adding an analogous
10\% uncertainty in the linear size and affects the $\chi^2$
minimization method making it increase after including QSO to
joint analysis. It means that even though the linear sizes
show negligible dependence on the luminosity and redshift,
uncertainty of the $l_m$ parameters can still be an important
source of systematic errors. In particular, the calibration
method for the linear size may also influence the goodness
of the final result.
Secondly, concerning the interaction term between dark energy
and dark matter, the statistical analysis based on three types of
standard probes demonstrates that the parameter describing the
interaction between DE and DM seems to be vanishing or slightly
smaller than zero. This conclusion is consistent with the previous
results for constraining IDE model parameters with other combined
observations. Thirdly, the estimated values of the Hubble constant
are in agreement with the standard ones reported by \emph{Planck}
\citep{Ade16a}. However, we still find strong degeneracies between
$\gamma_{i}$ and $H_0$, i.e., a negative value of interaction term
will lead to a larger value of the Hubble constant, which may
alleviate the tension of $H_0$ between the recent \emph{Planck} and
\emph{HST} measurements.

\begin{center}
 \begin{figure}[htbp]
\centering
 \includegraphics[angle=0,width=90mm]{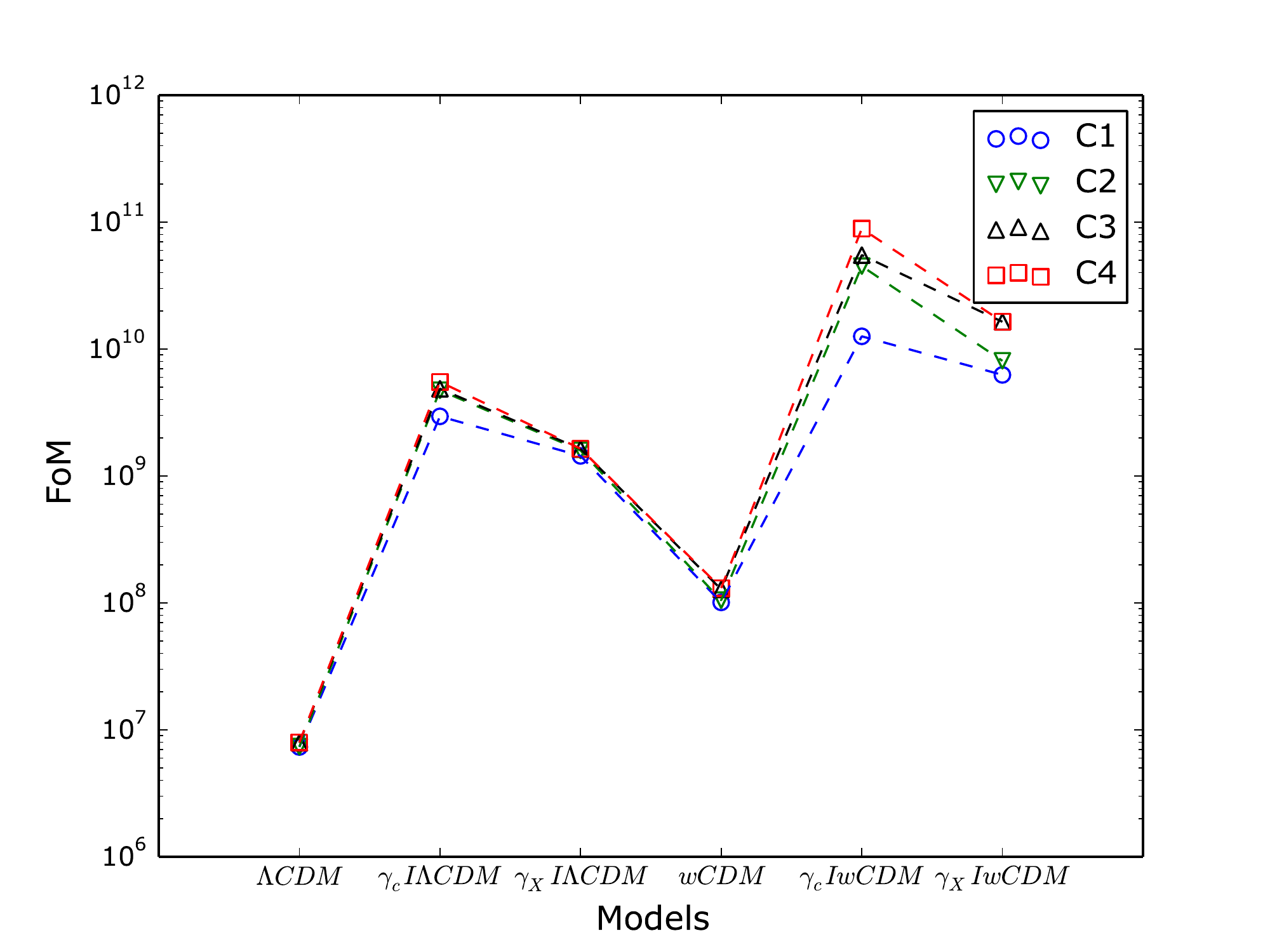}
 \caption{\label{figFoM} The FoM value for each cosmological model
 calculated from different data combinations:
 C1 (CMB+BAO), C2 (QSO+CMB+BAO), C3 (CMB+BAO+SN+$H(z)$) and
 C4 (QSO+CMB+BAO+SN+$H(z)$).}
 \end{figure}
 \end{center}

Finally, we would like to discuss statistically the performance of
our data sets and perform model selection. On the one hand, in order
to quantify the constraining power of the quasar data on
cosmological model parameters, we calculate the Figure of Merit
(FoM) for the IDE models with different data combinations. Based on
the the previous definition proposed by the Dark Energy Task Force
(DETF) for the CPL parametrization \citep{Albrecht06}, a more
general expression can be rewritten as \citep{Wang08}
\begin{equation}
FoM=(det Cov(\bf{p}))^{-1/2}
\end{equation}
where $Cov(\bf{p})$ is the covariance matrix of relevant
cosmological parameters $\bf{p}$. Higher value of the FoM is
corresponds to tighter model constraint. From the results
shown in Fig.~\ref{figFoM} and Table~\ref{tablea}, regarding the
comparison between statistical standard rulers (CMB+BAO)
and joint standard rulers (QSO+CMB+BAO), one can clearly see that
the inclusion of the quasars sample will generate more stringent
constraints on the $\gamma_c IwCDM$ model. Comparison
between the full combination (QSO+CMB+BAO+SN+$H(z)$) and the
combination excluding QSO data (CMB+BAO+SN+$H(z)$) concerning FoM leads to
similar conclusions. This can be understood in terms of
a higher redshift range covered by QSOs in comparison to other
astrophysical probes. However, for the rest of interacting dark
energy models considered in this paper, the significance of this
improvement is not evident and remains to be checked with a larger sample of quasars
observed by VLBI at high angular resolutions \citep{Pushkarev15}.

On the other hand, in the face of so many
competing interacting dark energy scenarios, it is important to find
an effective way to decide which one is most favored by the data.
Following the analysis of \citet{BiesiadaAIC} and \citet{Cao11b},
model comparison statistics including the Akaike Information
Criterion (AIC) \citep{Akaike74} and the Bayesian Information
Criterion (BIC) \citep{Schwarz78} will be used for this purpose. The
expression of the two main information criteria can be written as
\begin{equation}
AIC=-2\mathrm{ln}\mathcal{L}_{max}+2k=\chi^2_{min}+2k
\end{equation}
and
\begin{equation}
BIC=-2\mathrm{ln}\mathcal{L}_{max}+k\mathrm{ln}N=\chi^2_{min}+k\mathrm{ln}N
\end{equation}
where $\mathcal{L}_{max}=exp(-\chi^2_{min}/2)$ is the maximum
likelihood value, $k$ is the number of free parameters in the model
and $N$ is the total number of data points used in the statistical
analysis.

Table \ref{tableIC} lists the AIC and BIC difference of each model.
One can see that the two criteria lead to different conclusions,
concerning the ranking of competing dark energy models. More
specifically, according to the AIC, the $\gamma_c$I$\Lambda$CDM
performs the best, while the $\gamma_X$IwCDM model gets the smallest
support from the current observations. On the contrary, the BIC
criterion gives a different ranking: the cosmological constant model
is still the best cosmological model. Then comes the sequence: XCDM,
$\gamma_c$I$\Lambda$CDM, and $\gamma_X$I$\Lambda$CDM with one more
free parameter, while $\gamma_c$IwCDM and $\gamma_X$IwCDM are
clearly disfavored by the data. Therefore, our findings indicate
that interacting dark energy models with more free parameters are
substantially penalized by the BIC criterion, which agrees very well
with the previous results derived from other cosmological probes
including strong lensing systems \citep{Biesiada10,Cao11b,Li16}.

 \begin{table}[htp]
\centering \caption{ Summary of the information criteria for
different interacting dark energy scenarios, obtained from the
combinations of the data sets: C1 (CMB+BAO), C2 (QSO+CMB+BAO), C3
(CMB+BAO+SN+$H(z)$) and C4
(QSO+CMB+BAO+SN+$H(z)$). Corresponding results for the $\Lambda$CDM
and wCDM models are also added for comparison.}
{{\scriptsize
 \begin{tabular}{c  c  c  c c  c c } \hline\hline
Data  & $Model$  & $k$ & $AIC$ & $\Delta AIC$ & $BIC$ & $\Delta
BIC$\\ \cline{1-7}
          & $\Lambda CDM$           & $3$ & $16.35$ & $0$    & $18.04$ & $0$    \\
          & $\gamma_c I\Lambda CDM$ & $4$ & $16.95$ & $0.60$ & $19.21$ & $1.17$ \\
  $C1$    & $\gamma_X I\Lambda CDM$ & $4$ & $17.16$ & $0.81$ & $19.42$ & $1.38$ \\
$(N=13)$  & $wCDM$                  & $4$ & $16.86$ & $0.51$ & $19.12$ & $1.08$ \\
          & $\gamma_c IwCDM$        & $5$ & $18.76$ & $2.41$ & $21.58$ & $3.54$ \\
          & $\gamma_X IwCDM$        & $5$ & $18.85$ & $2.50$ & $21.67$ & $3.63$ \\ \cline{1-7}
          & $\Lambda CDM$           & $3$ & $370.55$ & $1.36$ & $379.22$ & $0$    \\
          & $\gamma_c I\Lambda CDM$ & $4$ & $369.19$ & $0$    & $380.75$ & $1.53$ \\
  $C2$    & $\gamma_X I\Lambda CDM$ & $4$ & $371.24$ & $2.05$ & $382.80$ & $3.58$ \\
$(N=133)$ & $wCDM$                  & $4$ & $371.29$ & $2.10$ & $382.85$ & $3.63$ \\
          & $\gamma_c IwCDM$        & $5$ & $371.19$ & $2.00$ & $385.64$ & $6.42$ \\
          & $\gamma_X IwCDM$        & $5$ & $373.13$ & $3.94$ & $387.58$ & $8.36$ \\ \cline{1-7}
          & $\Lambda CDM$           & $3$ & $577.26$ & $0.14$ & $590.56$ & $0$ \\
          & $\gamma_c I\Lambda CDM$ & $4$ & $577.12$ & $0$    & $594.86$ & $4.30$ \\
  $C3$    & $\gamma_X I\Lambda CDM$ & $4$ & $578.04$ & $0.92$ & $595.78$ & $5.22$ \\
$(N=623)$ & $wCDM$                  & $4$ & $578.96$ & $1.84$ & $596.70$ & $6.14$ \\
          & $\gamma_c IwCDM$        & $5$ & $579.07$ & $1.95$ & $601.24$ & $10.68$ \\
          & $\gamma_X IwCDM$        & $5$ & $579.97$ & $2.85$ & $602.14$ & $11.58$ \\ \cline{1-7}
          & $\Lambda CDM$           & $3$ & $931.36$ & $1.98$ & $945.19$ & $0$    \\
          & $\gamma_c I\Lambda CDM$ & $4$ & $929.38$ & $0$    & $947.82$ & $2.63$ \\
  $C4$    & $\gamma_X I\Lambda CDM$ & $4$ & $932.07$ & $2.15$ & $950.51$ & $5.32$ \\
$(N=743)$ & $wCDM$                  & $4$ & $933.13$ & $3.75$ & $951.57$ & $6.38$ \\
          & $\gamma_c IwCDM$        & $5$ & $931.16$ & $1.78$ & $954.21$ & $9.02$ \\
          & $\gamma_X IwCDM$        & $5$ & $933.94$ & $4.56$ & $956.99$ & $11.80$ \\
 \hline \hline
\end{tabular}} \label{tableIC}}
\end{table}

\section{Conclusions}\label{sec:conclu}

In this paper, using the newly released data, derived from the
very-long-baseline interferometry (VLBI) observations, comprising
redshifts and angular sizes of compact structure in radio quasars,
we have examined several popular phenomenological interaction models
for dark energy and dark matter. Such models were proposed to
alleviate the coincidence problem of the concordance $\Lambda$CDM
model as well the tension between $H_0$ derived from the recent
\emph{Planck} and \emph{HST} measurements. The sub-sample of 120
intermediate-luminosity quasars in the redshift range of
$0.46<z<2.8$, which was extracted from 613 milliarcsecond
ultra-compact radio sources observed by a 2.29 GHz VLBI all-sky
survey, can be used as cosmological standard rulers with the linear
sizes calibrated by a cosmological-model-independent method,
$l_m=11.03\pm0.25$ pc. Compared with the SN Ia standard candles
($z\sim1.4$) extensively used in the cosmological study, the
advantage of this data set is that quasars are observed at much
higher redshifts ($z\sim 3.0$), which motivate us to investigate the
possible interaction between dark energy and dark matter at higher
redshifts. To reduce the uncertainty and put tighter constraints on
the coupling parameters, we added, in our discussion, the CMB
results from the \emph{Planck} as well as the BAO measurements both
from the low-$z$ galaxy and higher-$z$ Lyman-$\alpha$ Forest
(Ly$\alpha$F) data. We expect that sensitivities of measurements of
different observables can give complementary results on the coupling
between dark sectors. Moreover, in order to quantify the
constraining power of the current quasar data on the model
parameters and the ranking of competing dark energy models, we
assessed the FoM and performed model comparison using
information-theoretical techniques (AIC and BIC criteria). Here we
summarize our main conclusions in more detail:
\begin{itemize}
  \item The quasar data could provide
constraints competitive to the other two types of standard probes,
i.e., the SNIa as standard candles with more precise data points and
cosmic chronometers providing $H(z)$ which are more directly related
to the cosmological model parameters of interest. Moreover, our
results strongly suggest that larger and more accurate sample of the
quasar data can become an important complementary probe to test the
properties of dark energy and break the degeneracy in the IDE model
parameters. This conclusion is strengthened by the statistical
results from the Figure of Merit, which supports the claim that the
inclusion of the QSO sample covering higher redshift range leads to
more stringent constraints on certain interacting dark energy models
(especially the $\gamma_c$IwCDM model).
\item The estimated values of the Hubble constant are in agreement with
the standard ones reported by \emph{Planck} \citep{Ade16a}. However,
strong degeneracies between $\gamma_{i}$ and $H_0$ implying that
negative value of the interaction term will lead to a larger value of
the Hubble constant, may alleviate the tension between the recent
\textit{Planck} and \emph{HST} measurements of $H_0$.
\item The AIC and BIC have provided quite different conclusions,
concerning the ranking of competing dark energy models.
More specifically, according to the AIC, the $\gamma_c$I$\Lambda$CDM
performs the best, while the $\gamma_X$IwCDM model gets the smallest
support from the current observations. On the contrary, the BIC
criterion gives a different ranking: the cosmological constant model
is still the best cosmological model followed by the $w$CDM,
$\gamma_c$I$\Lambda$CDM, and $\gamma_X$I$\Lambda$CDM, while
$\gamma_c$IwCDM and $\gamma_X$IwCDM are clearly disfavored by the
data. Therefore, our findings indicate that interacting dark energy
models with more free parameters are substantially penalized by the
BIC criterion, which agrees very well with the previous results
derived from other cosmological probes.
\end{itemize}

\section{Acknowledgments}

This work was supported by the National Key Research and Development
Program of China under Grants No. 2017YFA0402603; the Ministry of
Science and Technology National Basic Science Program (Project 973)
under Grants No. 2014CB845806; the National Natural Science
Foundation of China under Grants Nos. 11503001, 11373014, and
11690023; the Fundamental Research Funds for the Central
Universities and Scientific Research Foundation of Beijing Normal
University; China Postdoctoral Science Foundation under grant No.
2015T80052; and the Opening Project of Key Laboratory of
Computational Astrophysics, National Astronomical Observatories,
Chinese Academy of Sciences. X.Z. is supported by the China
Scholarship Council. M.B. gratefully acknowledges support and
hospitality of the Beijing Normal University.

%
%

\end{document}